\documentclass[sn-nature]{sn-jnl}

\usepackage{graphicx}%
\usepackage{multirow}%
\usepackage{amsmath,amssymb,amsfonts}%
\usepackage{amsthm}%
\usepackage{mathrsfs}%
\usepackage[title]{appendix}%
\usepackage{xcolor}%
\usepackage{textcomp}%
\usepackage{manyfoot}%
\usepackage{booktabs}%
\usepackage{algorithm}%
\usepackage{algorithmicx}%
\usepackage{algpseudocode}%
\usepackage{listings}%

\theoremstyle{thmstyleone}%
\theoremstyle{thmstyletwo}%

\theoremstyle{thmstylethree}%

\begin{document}

\title{A performance evaluation of integrating machine learning schemes utilizing fluidic lenses}

\author[1,2]{ Graciana Puentes}


\affil[1]{Departamento de Fisica, Facultad de Ciencias Exactas y Naturales, Universidad de Buenos Aires, Ciudad Universitaria, 1428 Buenos Aires, Argentina}

\affil[2]{CONICET-Universidad de Buenos Aires, Instituto de Fisica de Buenos Aires (IFIBA), Ciudad Universitaria, 1428 Buenos Aires, Argentina}


\abstract{ A combination of statistical inference and machine learning (ML) schemes has been utilized to create a thorough understanding of coarse experimental data based on Zernike variables characterizing  optical aberrations in fluidic lenses. A classification of surplus-response variables through tolerance manipulation was included to unravel the dimensional aspect of the data. Similarly, the impact of the exclusion of supererogatory variables through the identification of clustering movements of constituents is examined. The method of constructing a spectrum of collaborative results through the application of similar techniques has been tested. To evaluate the suitability of each statistical method before its application on a large dataset, a selection of ML schemes has been proposed. The supervised learning tools principal component analysis (PCA), factor analysis (FA), and hierarchical clustering (HC) were employed to define the elemental characteristics of Zernike variables. PCA enabled to reduce the dimensionality of the system by identifying two principal components which collectively account for 95\% of the total variance. The execution of FA indicated that a specific tolerance of independent variability of 0.005  could be used to reduce the dimensionality of the system without losing essential data information. A high cophenetic coefficient value of c=0.9629 validated an accurate clustering division of variables with similar characteristics. The current approach of mutually validating ML and statistical analysis methods will aid in laying the foundation for state-of-the-art (SOTA) analysis. The benefit of our approach can be assessed by considering that the associated SOTA will enhance the predictive accuracy between two comparable methods, in contrast to the SOTA analysis conducted between two arbitrary ML methods. 
}

\maketitle


\section{Introduction}

In recent years numerous academic initiatives have been undertaken to evaluate coarse data quality through the application of machine learning (ML) and artificial intelligence (AI) techniques \cite{Zhao, Cirrincione, Zhang, Guan, Du, Jain, Mangiameli, Herrero, Okamoto, Shahid, Shahid2}.  It has been suggested that the classification of self-organizing maps (SOM) could effectively represent the overall data quality within a distribution network in a direct and vivid manner by projecting high-dimensional data quality indicators onto a low-dimensional topology grid. Furthermore, it has been determined that a two-stage classification approach utilising SOM alongside clustering methods yielded greater efficiency when compared to traditional clustering techniques.  Three ML models, namely boosted regression trees (BRT), multivariate discriminant analysis (MDA), and support vector machine (SVM), were utilized to create an innovative framework for evaluating the risk of contamination in environmental data \cite{Sajedi}. The accuracy of these models was established leading to their consideration for an ensemble modelling process. Furthermore,  significant scholarly efforts have been made to assess coarse data using ML and AI schemes using SOM and clustering methods, resulting in  higher efficiency in comparison to the traditional clustering methods \cite{Zhu}.
The applicability of ML algorithms in assessing trends in medical, agricultural, financial, environmental and sports data, as well as in enhancing model predictability, has been demonstrated revealing the potentiality of statistical methods to analyze a monumental multi-parameters data with error precision \cite{Ibrahim, Lopez}.  A new classification of ML algorithms was introduced to emphasize the significance of developing refined algorithms to address challenges such as topology alterations, link failures, memory limitations, interoperability, network congestion, coverage, scalability, network management, security, and privacy \cite{Maione}. In light of these computational explorations aimed at evaluating the correlation of data parameters and their potential impact on the system, we propose to utilize a combination of ML algorithms and statistical inference tools to study an ensemble of Zernike variables characterizing optical aberrations in fluidic lenses.\\

\begin{figure}
\centering
\includegraphics[width=1\linewidth]{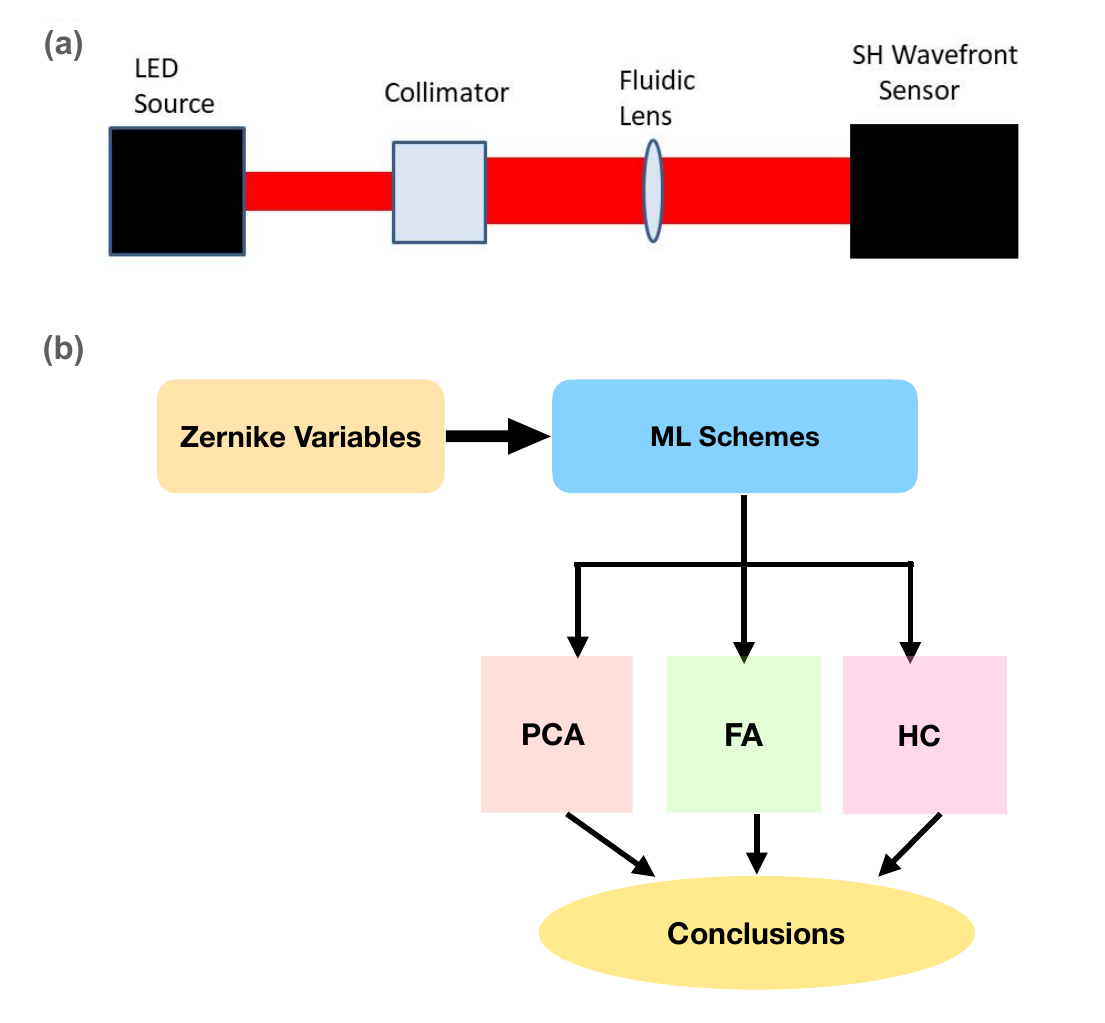}
\caption{\label{fig:frog} (a) Experimental setup for reconstruction of Zernike coefficients characterizing the phase front transmitted by fluidic lenses of different fluidic volumes, using a Shack-Hartmann (SH) wave-front sensor. (b) Study flow chart. }
\end{figure}

Specifically, the present article offers a synthesis of statistical analysis and ML tools to achieve a profound understanding of coarse data. Multi-parameter data concerning Zernike components characterizing optical aberrations in the phase-front transmitted by fluidic lenses of different fluidic volumes, obtained using a Shack-Hartmann wave-front sensor reconstruction method \cite{1,2,3,4,5,6,7,8,9,10,10b}, has been utilized for experimentation. The study's objectives are as follows: (i) An analysis of the data will be conducted to fulfill a dual purpose: evaluating Zernike variable quality and demonstrating the discerning ML techniques, supported by multivariable statistical inference. (ii) The appropriateness and effectiveness of each computational method will be assessed before its application to the data. (iii) Considering the probabilistic inferences derived from statistical methods and their integration into the ML decision-making process, a range of collaborative results will be developed to advance the Zernike variable  valuation process. The quantitative and descriptive outcomes of each method will be contemplated to emphasize the key features of the data. (iv) An aspect of eliminating redundant variables will be integrated into the analysis by identifying the clustering behavior of similar characteristic variables. (v) A variety of data visualization techniques will be employed to present the results obtained from the combination of ML and statistical methods. (vi) A novel approach for mutually validating ML and statistical analysis  techniques will be introduced to lay the foundation for state-of-the-art (SOTA) analysis. The benefits of our approach can be illustrated by the fact that the related SOTA analysis will enhance the predictive accuracy between two comparable ML methods, in contrast to the SOTA between two arbitrary ML methods.

\section{Data Components}

The data used for the current analysis is comprised of 15 Zernike coefficients for 8 different weights, obtained by using a Shack-Hartmann Wavefront Sensor to reconstruct the phase-front transmitted by fluidic lenses of varying fluidic volumes \cite{1,2,3,4,5,6,7,8,9,10,11,12,13,14,15,16,17,18,19,20,21,22,23}, as depicted in Figure 1 (a). The 15 Zernike coefficients ($Z_1-Z_{15}$) and their associated Zernike polynomial for characterisation of optical aberrations are depicted in Table 1 left column and right column, respectively. The number assigned to the Zernike variables depicted in Table 1 (middle column) correspond to the numbering system used for HC analysis.

\begin{table}[h]
\begin{tabular}{@{}llll@{}}
\toprule
\textbf{Zernike coefficient}  & \textbf{Nr Assigned to Variable} & \textbf{Zernike polynomial}  \\
\midrule
$Z_{0}$ & 1 &1 \\
$Z_{1}$ & 2 & $2\xi \sin \phi$  \\
$Z_{2}$ & 3 & $2\xi \cos \phi$ \\
$Z_{3}$ & 4 & $\sqrt{6}\xi ^{2}\sin 2\phi $ \\
$Z_{4}$ & 5 & $\sqrt{3}\left( 2\xi ^{2}-1\right) $ \\
$Z_{5}$ & 6 &$ \sqrt{6}\xi ^{2}\cos 2\phi $\\
$Z_{6}$ & 7 & $\sqrt{8}\xi ^{3}\sin 3\phi$\\
$Z_{7}$ & 8 &$ \sqrt{8}\left( 3\xi ^{3}-2\xi \right) \sin \phi $\\
$Z_{8}$ & 9 & $\sqrt{8}\left( 3\xi ^{3}-2\xi \right) \cos \phi $\\
$Z_{9}$ & 10 &$ \sqrt{8}\xi ^{3}\cos 3\phi$ \\
$Z_{10}$ & 11 & $\sqrt{10}\xi ^{4}\sin 4\phi $\\
$Z_{11}$ & 12 &$ \sqrt{10}\left( 4\xi ^{4}-3\xi ^{2}\right) \sin 2\phi $ \\
$Z_{12}$ & 13 &$ \sqrt{5}\left( 6\xi ^{4}-6\xi ^{2}+1\right) $\\
$Z_{13}$ & 14 & $\sqrt{10}\left( 4\xi ^{4}-3\xi ^{2}\right) \cos 2\phi $ \\
$Z_{14}$ & 15 & $\sqrt{10}\xi ^{4}\cos 4\phi $\\
\botrule
\end{tabular}
\caption{\label{tab:widgets}Left Column: Zernike coefficients obtained by wavefront sensing ($Z_1-Z_{15}$). Right column: Associated Zernike polynomial. Middle column: Number Assigned to Variables $Z_1-Z_{15}$ for HC and KMC analysis. }
\end{table}

\begin{figure}[h!]
\centering
\includegraphics[width=0.9\linewidth]{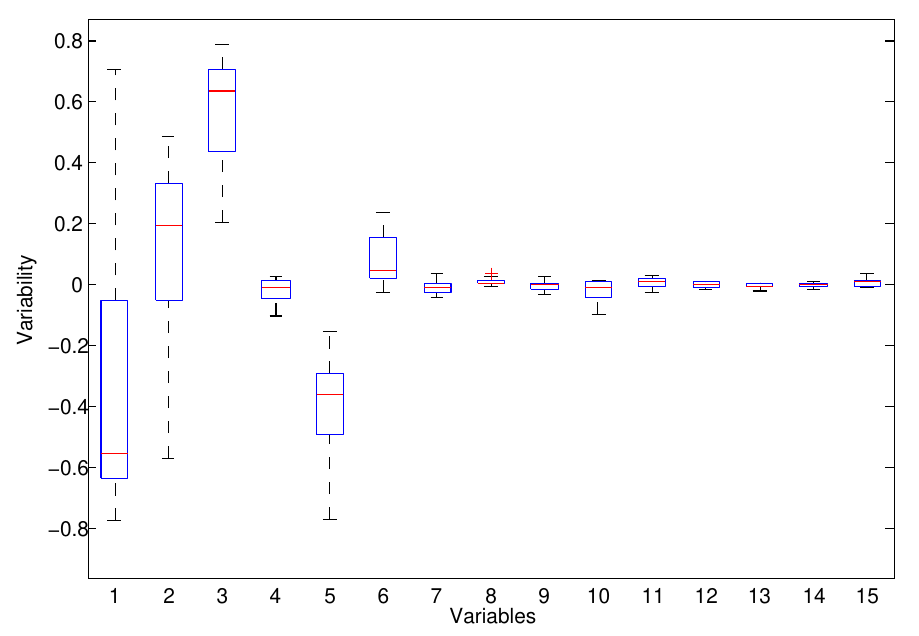}
\caption{\label{fig:frog} Boxplot displaying the variability of Zernike variables. }
\end{figure}

\begin{figure}[h!]
\includegraphics[width=1\linewidth]{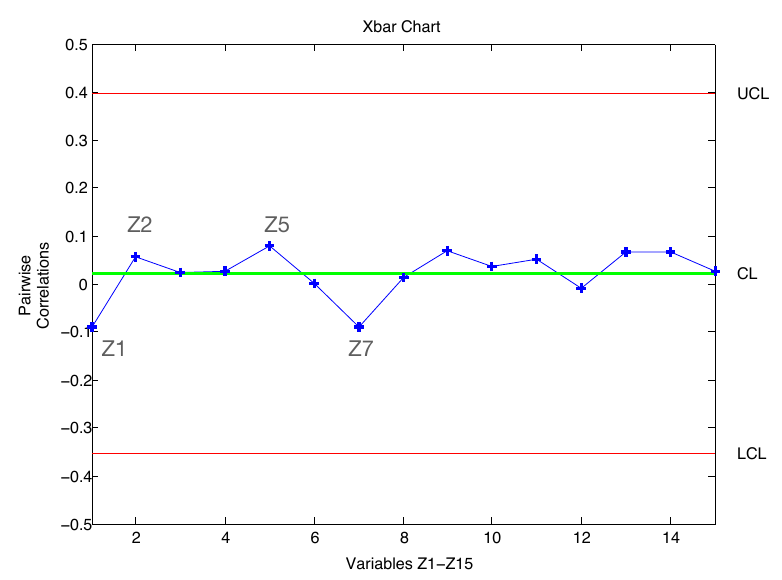}
\caption{\label{fig:frog} XBAR control chart displaying pairwise correlations for Zernike variables $Z_1$-$Z_{15}$. }
\end{figure}

\begin{figure}[h!]
\includegraphics[width=1\linewidth]{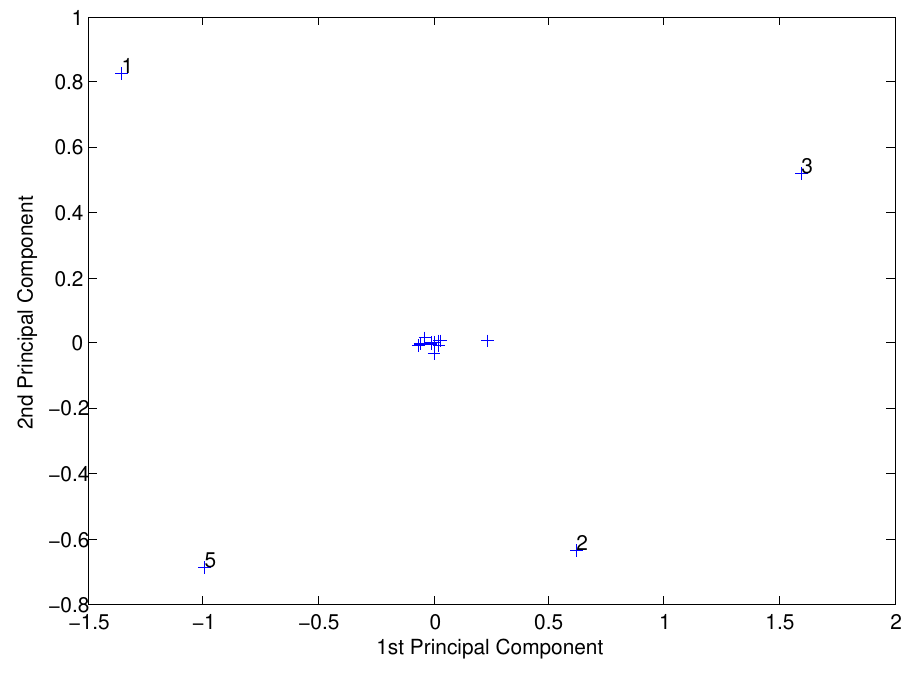}
\caption{\label{fig:frog} Principal components plot. }
\end{figure}

\begin{figure}[t!]
\includegraphics[width=1\linewidth]{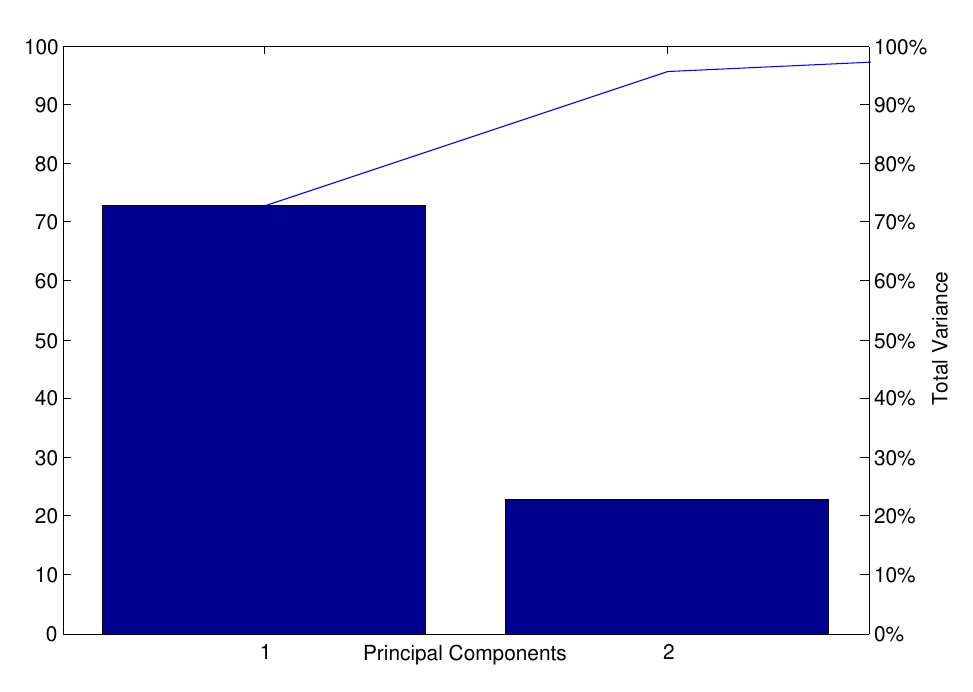}
\caption{\label{fig:frog} Scree plot. }
\end{figure}

\begin{figure}[t!]
\includegraphics[width=1\linewidth]{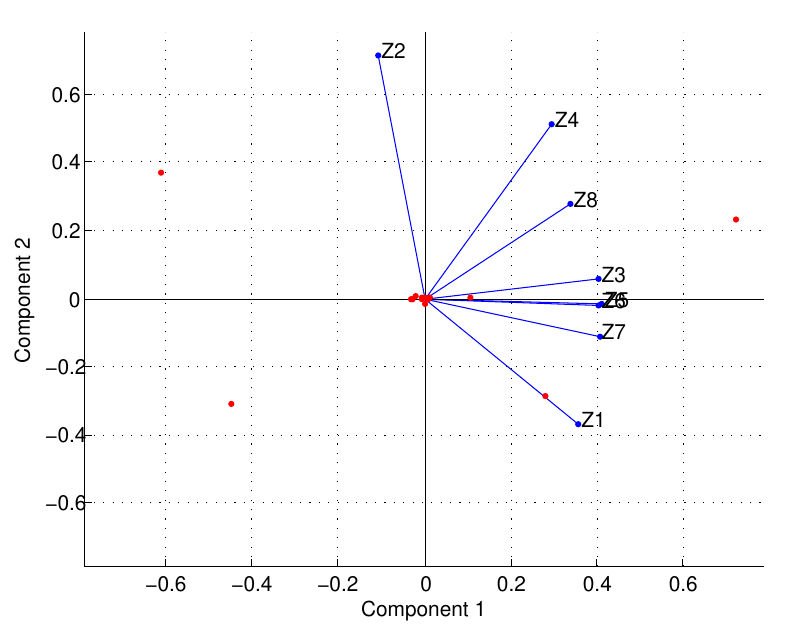}
\caption{\label{fig:frog} Bi-plot displaying principal components coefficients and principal components scores (red dots). }
\end{figure}

\begin{figure}[h!]
\includegraphics[width=1\linewidth]{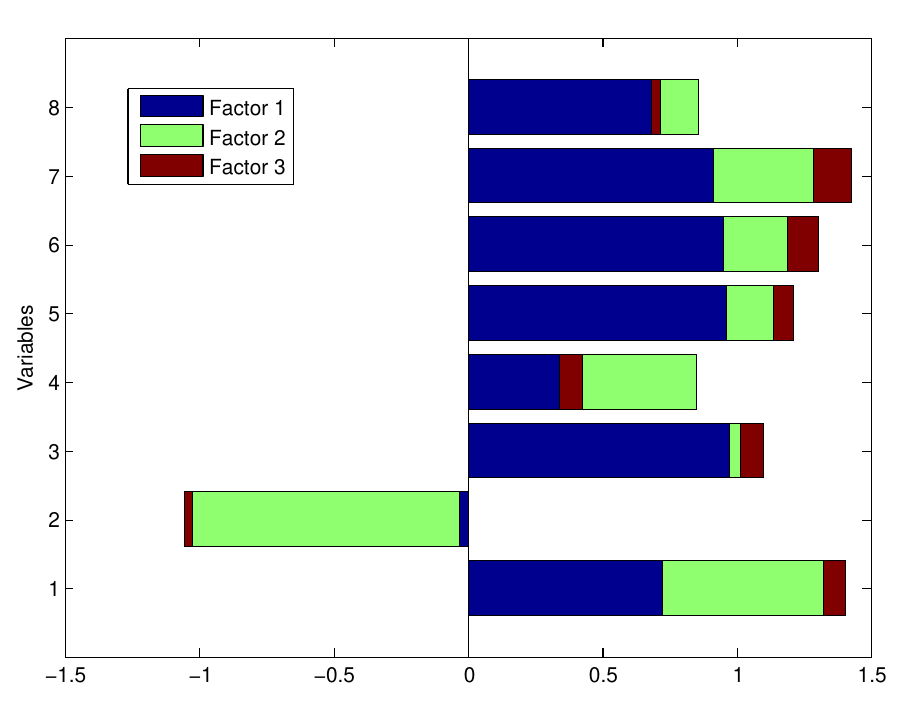}
\caption{\label{fig:frog} Rotated estimated loadings for Factor Analysis. }
\end{figure}

\begin{figure}[h!]
\includegraphics[width=1\linewidth]{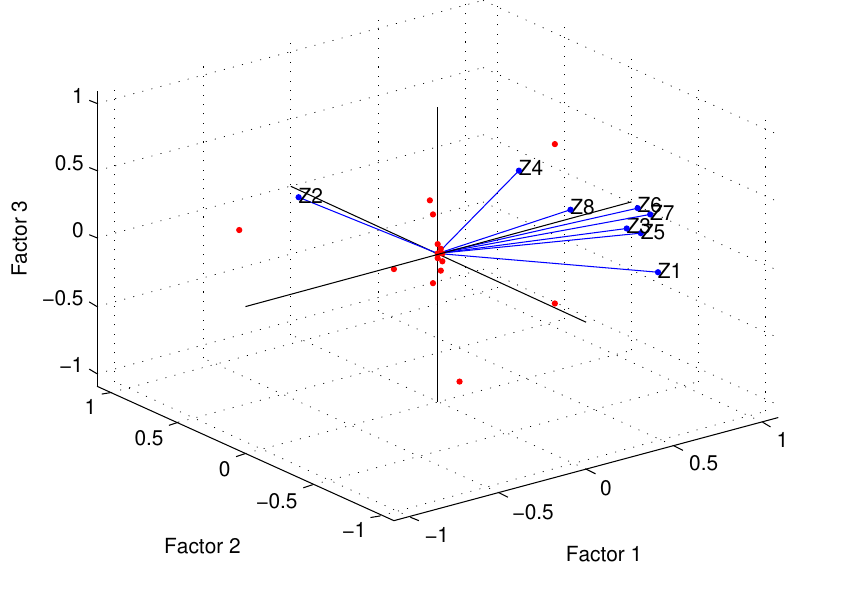}
\caption{\label{fig:frog} Rotated factors (1-3) score plot. }
\end{figure}


The experimental setup (Fig. 1 (a)) used to obtain the raw data analyzed by the different ML schemes consists of the following elements: A collimated incoherent beam, produced by an LED source ($\lambda$ = 633 nm) propagates through a fluidic lens, consisting of an elastic PDMS double membrane filled with a fluid of matching index of refraction and variable volume. The phase-front transmitted by the fluidic lens is imaged by a Schack-Hartmann wave-front sensor (Model Thorlabs WFS150-5C), located at a distance of 2 cm from the fluidic lens, in order to image the near field produced by the lens. The area of the beam to be characterized is determined by the aperture of the sensor (typically 3mm). We measured the optical aberrations in the wave-front transmitted by the fluidic lens in terms of the coefficients associated with the first 15 Zernike polynomials for 8 different fluidic volumes  \cite{7,9,10,10b}. The 15 Zernike coefficients for 8 different fluidic volumes are used as input variables and weights for statistical and ML analysis.

\section{Statistical Methods}


\subsection{Boxplot}

A boxplot of the normalized data has been created to provide insights into significant information regarding the variability of Zernike variables. The boxplot displayed in Fig. 2 reveals the constancy of Zernike variables $Z_7-Z_{15}$ for several different volumes, in addition to displaying the higher variability of $Z_1$ compared to $Z_2-Z_6$. This information was used  as a criteria to organize and reduce the dimensionality of the Zernike data.

\subsection{XBAR control chart}

An assessment of pairwise correlations between each pair of columns among the variables in the dataset has been conducted. It has been noted that certain variables exhibit a higher correlation in comparison to other variables. The XBAR control chart in Fig. 3 displays the pairwise correlations of different Zernike variables. XBAR chart plots the arithmetic mean (CL) of each subgroup, and indicates out-of-control points that are more than three standard deviations above (UCL) and below (LCL) the mean. Highly correlated variables above zero are $Z_2, Z_5$ and below zero are $Z_1, Z_7$, while least correlated variables are $Z_3$, $Z_4$, $Z_6$ and $Z_8-Z_{15}$. No violation points among pairwise correlations were identify.



\subsection{Principal Components Analysis}

In principal component analysis (PCA), an original set of variables is transformed into a new set of variables, referred to as principal components (PC). Each PC is derived from a linear combination of the original variables. These PCs serve as an orthogonal basis for the entire dataset. We conducted PCA using the inverse variances of Zernike variables as weights. The analysis yielded five key outputs (coefficients, scores, latent, T-squared, explained), which we summarize here:

The first output presents the weighted coefficients of the PC derived from the original variables, which are then adjusted to create an orthonormal basis. The second output, known as the ‘component score’ , reflects the coordinates of the original data within a new coordinate system established by the PCs. In the component plot (Fig. 4), we can see the centred and scaled data projected onto the first two PCs. Notably, the Zernike coefficient $Z_1, Z_2, Z_3 , Z_5$ display a pronounced deviation compared to other Zernike variables.

The third output, termed ‘latent,’ calculates the variances associated with each PC. Each score column corresponds to a sample variance aligned with the respective row of latent. To assess the multivariate distance of each observation from the data center, we computed the fourth output, T-squared or Hotelling’s $T^2$. This also serves as a method to identify extreme data points. The fifth and final output, labeled ‘explained,’ includes the percentage of variances accounted for by each principal component. 

The Scree plot (Fig. 5) illustrates the explained variance for each principal component, showing that the first two PCs collectively explain 95\% of the total variance. Specifically, the first PC alone accounts for 70\% of the variance, indicating that additional components may be necessary. Ideally, in PCA the first few components should explain a large percentage of the total variance. If the first two  components explain most of the variance, this means that the data is well-represented in the lower-dimensional space.

Based on the numerical data obtained from the PCA, we created a bi-plot (Fig. 6) that simultaneously displays the orthonormal principal component coefficients for each variable along with the principal component scores (red dots) for each observation. This visualization reveals that the variables $Z_2$  has the most significant contributions to the first two principal components, exhibiting the largest coefficient values.  This graph also points to the fact that first and second principal component classifies  variables with positive and negative coefficients, where $Z_3,Z_4,Z_8$ have a positive 1st and 2nd principal coefficients. We emphasize that this result was obtained after normalizing the  Zernike data. A different PC classification can be obtained for unnormalized Zernike variables. 

The data presented in this Fig. 6 can be used to segment large datasets into smaller groups in order to examine the impact of their total variability on the overall system. The application of PCA to analyze Zernike coefficient data was found to yield precise and meaningful results. Additionally, incorporating ML-PCA into our analysis has enhanced the computational efficiency of the task, as the algorithm simultaneously considered weighted coefficients and components of latent and T-squared elements.

\subsection{Factor Analysis}

\begin{table}
\centering
\begin{tabular}{|c|c|c|c|}
\hline
 \textbf{Variable}  & \textbf{Factor 1} & \textbf{Factor 2} & \textbf{Factor 3} \\
 \hline
$Z_1$ & 0.9072 & -0.4015  & 0.0927\\
$Z_2$ & -0.3613 & 0.9288  & 0.0476\\
$Z_3$ & 0.9566 & 0.1978  & 0.1190\\
$Z_4$ & 0.6343 & 0.7695  & -0.0455\\
$Z_5$ & 0.9888 & 0.0863  & 0.1006\\
$Z_6$ & 0.9768 & 0.0966  & -0.0899\\
$Z_7$ & 0.9901 & -0.0418  & -0.1158\\
$Z_8$ & 0.7607 & 0.4172  & 0.0693\\
\hline 
\end{tabular}
\caption{\label{tab:widgets} Loading matrix for factor analysis}
\end{table}

\begin{table}
\centering
\begin{tabular}{|c|c|c|c|c|c|c|c|c|}
\hline
 \textbf{Variable}  & \textbf{Z1} & \textbf{Z2} & \textbf{Z3} & \textbf{Z4} & \textbf{Z5} & \textbf{Z6} & \textbf{Z7} & \textbf{Z8} \\
 \hline
Variance & 0.0073 & 0.0050  & 0.0315 & 0.0050 & 0.0050 & 0.0285 & 0.0050 & 0.2427\\
\hline 
\end{tabular}
\caption{\label{tab:widgets}Specific variance for factor analysis}
\end{table}

\begin{table}
\centering
\begin{tabular}{|c|c|c|c|}
\hline
 \textbf{Variable}  & \textbf{Factor 1} & \textbf{Factor 2} & \textbf{Factor 3}  \\
 \hline
$Z_1$ &  \textbf{0.7246} & 0.6789  & -0.0522\\
$Z_2$ & -0.0298 &  \textbf{-0.9969}  & -0.0267\\
$Z_3$ &  \textbf{0.9715} & 0.1300  & -0.0877\\
$Z_4$ &  \textbf{0.8521} & -0.5126  & 0.0870\\
$Z_5$ &  \textbf{0.9640} & 0.2463  & -0.0720\\
$Z_6$ & \textbf{0.9495} & 0.2368  & 0.1183\\
$Z_7$ & \textbf{0.9151} & 0.3722  & 0.1400\\
$Z_8$ & \textbf{0.8580} & -0.1410  & -0.0349\\
\hline 
\end{tabular}
\caption{\label{tab:widgets}Loading matrix for factor analysis in rotated coordinate system highlighting highest rotated loading values.}
\end{table}

Factor analysis (FA) is a ML scheme to fit a model to multivariate data to estimate interdependence. In a FA model, the measured variables depend on a smaller number of unobserved (latent) factors. Because each factor might affect several variables in common, they are known as common factors. Each variable is assumed to be dependent on a linear combination of the common factors, and the coefficients are known as loadings. Each measured variable also includes a component due to independent random variability, known as specific variance because it is specific to one variable.

FA inspection of Zernike data was performed using the ML ‘factoran' function, which returns the maximum likelihood estimate (MLE) of the factor loadings matrix, in a common factor analysis model with $m$ common factors. FA has provided significant insights into the interdependence of variable groups, which can be utilised to pair specific components for examining their combined effects. The results of FA for $m=3$ (Table 3), derived from the ML computational program 'factoran', have led to the elimination of certain variables with specific variance below 0.005. It is evident from the variance figures that only the variable $Z_8$ exhibits a higher level of independent variability, specifically 0.2427. Positive variances together with a positive error degrees of freedom value (dfe=7) asserted the use of $m=3$ common factors in making an accurate analysis.  

Another important outcome of FA, the rotated estimates of loadings (coefficients of variables represented in the three common factors), identified which variables were primarily characterized by each of the three factors. As illustrated in Fig. 7, the variables $Z_1-Z_8$ are associated with the first (blue), second (green), third (brown) factors, respectively. This observation is further supported by the highest rotated loading values (highlighted in Table 4) of the variables in relation to their corresponding factors. Additionally, to reinforce the idea that each variable is influenced by only one common factor, plots of rotated factor scores and rotated factor loadings were generated. Fig. 8  clearly illustrates the variables ($Z_1-Z_8$) aligned with their respective factors along with their associated coefficient values. Remarkably, rotated Factor 3 only returns positive loading components. 


\subsection{Hierarchical Clustering}

To identify variables with shared attributes within the Zernike coefficient data, a Hierarchical Clustering (HC) method was utilized \cite{37}. This technique organizes data across various scales by constructing a cluster tree or dendrogram. Instead of offering a single set of clusters, the tree represents a multilevel hierarchy where clusters at one level are merged into larger groups at the next. This hierarchical structure allows for flexibility in determining the optimal level or scale of clustering, facilitating the analysis of collective movements or the impact of specific clusters within a system. The HC methodology was applied to 15 Zernike input variables, with the assigned numbers for these variables in the HC analysis displayed in Table 1 (middle column). It is important to highlight that HC was performed using normalized Zernike data to maintain consistency with the PCA and FA analyses conducted. Using unnormalized data would yield different clustering results \cite{10b}.


\begin{table}[h]
\begin{tiny}
\begin{tabular}{|l|l|l|l|l|l|l|l|l|l|l|l|l|l|l|}
\hline
\textbf{Sr} & 1& 2 &3 & 4 & 5 & 6 & 7 &8 & 9 &10 &11 & 12 & 13 &14  \\ \hline %
\textbf{Links} & 1 & 2 & 2 & 2 & 2 & 2 & 2 & 1 & 3 & 2 & 2 & 2 & 2 & 2  \\ \hline 
\textbf{Inconsistency} & 0 & 0.707 & 0.707 & 0.707 & 0.707 &0.707 & 0.707 & 0 & 1.054 & 0.707  & 0.707 & 0.707  & 0.707 & 0.707   \\ \hline 
\end{tabular}
\caption{\label{tab:widgets}Inconsistency measure for links in Dendrogram.}
\end{tiny}
\end{table}

\begin{figure}
\hspace{0.5cm}
\includegraphics[width=0.9\linewidth]{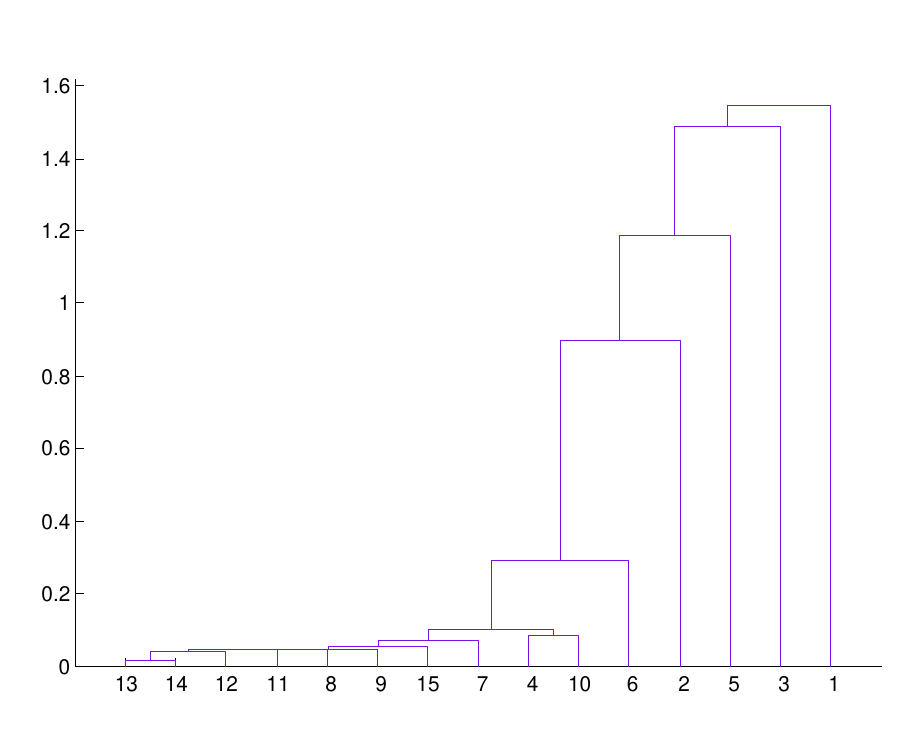}
\caption{\label{fig:frog} Dendrogam for clusters using normalized Zernike variables. }
\end{figure}

\begin{figure}
\hspace{0.5cm}
\includegraphics[width=0.9\linewidth]{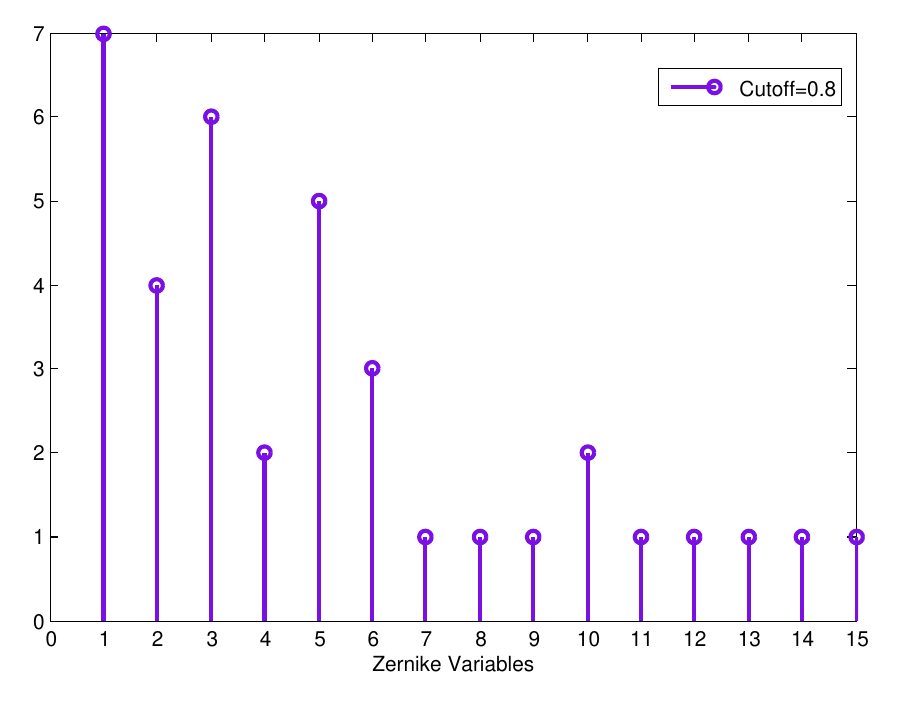}
\caption{\label{fig:frog} Stem plot setting an inconsistency cutoff of 0.8, resulting in the formation of 7 clusters.}
\end{figure}

To classify 15 Zernike variables into distinct groups, a cluster tree was developed using the ML program ‘clusterdata'. This hierarchical tree organizes clusters across multiple levels, grouping variables based on their comparable impact on the overall system. By applying the clusterdata function to the normalized dataset of 15 variables, a dendrogram (Fig. 9) was created utilizing a dissimilarity matrix and a linkage matrix. The dissimilarity matrix measured the distance between each variable pair, while the linkage matrix determined the connections between pairs of variables or formed clusters. Additionally, the linkage function advanced the clustering process by computing distances not only between individual variables but also between clusters or between a variable and a cluster, thereby refining the analysis further.

To validate the formation and connectivity of clusters represented in the cluster tree shown in Fig. 9, the cophenetic coefficient is calculated, resulting in a numerical value of $c=0.9629$. This coefficient serves as a key metric for assessing the correlation between the original distance matrix and the linkage matrix. Higher values, approaching 1, suggest that the distances between clusters or variables merged by the linkage method closely correspond to the actual distances derived from the dissimilarity function. Therefore, the cophenetic correlation coefficient for the dendrogram confirms that the linked clusters or variables accurately reflect their true spatial relationships within the input data. We stress that the value of $c$ obtained in the current analysis corresponds to normalized Zernike data. A different cophenetic coefficient can be obtained for unnormalized data  \cite{10b}.

To group a set of variables into natural clusters, an inconsistency measure is used to assess their similarities. The smallest value of this metric between any two variables or clusters reflects a close resemblance or strong similarity between them. This measure helps identify clusters based on the heights of the connections linking them. Table 5 outlines the inconsistency measures for the links represented in the dendrogram shown in Fig. 9. The table's second row details the number of links at various levels, while the third row presents their corresponding inconsistency coefficients. A Stem plot displaying clustering segmentation considering an inconsistency cutoff of 0.8 is presented in Fig. 10. The elected cutoff results in the formation of 7 distinct clusters. 

\section{Discussion}

The following limitations of the current study and potential directions for future research have been identified:

\begin{itemize}
\item{The applied ML algorithms require a robust, high-performance, computing infrastructure to operate effectively.}

\item{Prior to applying statistical and ML approaches, Zernike components ensemble underwent organization and normalization processes. Using these algorithms on unnormalized data could produce different outcomes.}

\item{A dimensionality reduction factor was incorporated into the study due to the negligible variability of certain parameters. As a result, predictive accuracy of the ML models is unlikely to reach a perfect hundred percent.}

\item{Expanding the scope of this analysis by including additional parametric components and a larger number of samples could enhance the comprehensiveness of the results.}

\item{An ensemble learning approach can be utilized to evaluate Zernike variables, providing a basis for comparison with existing results to assess accuracy.}

\end{itemize}

There is significant overlap in terminology and goals between PCA and FA. Much of the literature on the two methods does not distinguish between them, and some algorithms for fitting the FA model involve PCA. Both are dimension-reduction techniques, in the sense that they can be used to replace a large set of  observed variables with a smaller set of new variables. They also often give similar results. However, the two methods are different in their goals and in their underlying models. Roughly speaking, PCA should be used in order to approximate your data using fewer dimensions, and FA should be used when a explanatory model for correlations among the data is required. 

\section{Conclusions}

The application of the Machine Learning (ML) scheme principal component analysis (PCA) using weighted variables not only identifies critical samples among 15 Zernike variables and 8 distinct volumes but also categorises Zernike coefficients aligned with PCA components. Integrating ML-PCA insights with an unsupervised ML clustering method enabled precise parameter centralization and dimensionality reduction,  concluding that the first two principal components collectively explain 95\% of the data variability. The implementation of Factor Analysis (FA) through the ML program 'factoran' demonstrated that a specific tolerance for independent variability (0.005) can effectively reduce system dimensionality to a maximum of eight Zernike variables without compromising fundamental data integrity.  Additionally, a hierarchical clustering (HC) method generated a dendrogram to group variables with similar characteristics, validated by a high cophenetic coefficient ($c = 0.9629$), ensuring accurate cluster division. We stress that the current analysis was applied to normalized Zernike variables, different clustering configurations can be obtained utilizing unnormalized data \cite{10b}. Using the ML Stem algorithm, an inconsistency cutoff of 0.8 was elected for efficiently segmenting the large set of Zernike variables into seven distinct clusters, based on similarity and consistency criteria. The integration of insights from prior ML algorithms into subsequent assessments showcases improved distribution understanding, clustering divisions, and sample categorisations based on outlier components, all achieved with enhanced computational precision.

\section*{Funding} 

The author acknowledges financial support from FONCyT (PICT Startup 2015 0710). 

\section*{Disclosures}

The author declares no conflicts of interest.  

\section*{Data Availability}

Data underlying the results presented in this paper may be obtained from the author upon reasonable request.


\begin{thebibliography}{}

\bibitem{Zhao} Y. Zhao, G. Karypis, and U. Fayyad, "Hierarchical clustering algorithms for document datasets,"  Data Min. Knowl. Disc. \textbf{10}, 141–168 (2005).

\bibitem{Cirrincione} G. Cirrincione, G. Ciravegna, P. Barbiero, V. Randazzo, and E. Pasero, "The GH-EXIN neural network for hierarchical clustering,"  Neural Netw. \textbf{121}, 57–73 (2020).

\bibitem{Zhang} Z. Zhang, F. Murtagh, S. Van-Poucke, S. Lin, and P. Lan, "Hierarchical cluster analysis in clinical research with heterogeneous study population: Highlighting its visualization with R,"  Ann. Transl. Med. \textbf{5}, 75-86 (2017).

\bibitem{Guan} C. Guan and K. K. F.  Yuen, "The cognitive comparison enhanced hierarchical clustering,"  Granul. Comput. \textbf{7}, 637–655 (2022).

\bibitem{Du} K.-L. Du, "Clustering: A neural network approach,"  Neural Netw. \textbf{23}, 89–107 (2010).

\bibitem{Jain} A. Jain, M.  Murty, and P. Flynn, "Data clustering: A review,"  ACM Comput. Surv. \textbf{31}, 264–323 (1999).

\bibitem{Mangiameli} P. Mangiameli, S. K. Chen, and  D.  West, "A comparison of SOM neural network and hierarchical clustering methods,"  Eur. J. Oper. Res. \textbf{93}, 402–417 (1996).

\bibitem{Herrero} J. Herrero, A. Valencia, and J. Dopazo,  "A hierarchical unsupervised growing neural network for clustering gene expression patterns,"  Bioinformatics \textbf{17}, 126–136 (2001).

\bibitem{Okamoto} M. Okamoto, N. Bu, and  T.  Tsuji, "Unsupervised learning for hierarchical clustering using statistical information,"  Advances in Neural Networks 834–839 (2004).

\bibitem{Shahid} N. Shahid, "Comparison of hierarchical clustering and neural network clustering: an analysis on precision dominance,"  Sci. Rep. \textbf{13}, 5661-5672 (2023).

\bibitem{Shahid2} N. Shahid, "A proficiency assessment of integrating machine learning (ML) schemes on Lahore water ensemble,"  Sci. Rep. \textbf{13}, 5130-5149 (2023).

\bibitem{Sajedi} Sajedi-Hosseini, F. et al., "A novel machine learning-based approach for the risk assessment of nitrate groundwater contamination,"
Sci. Total Environ. \textbf{644}, 954–962 (2018).

\bibitem{Zhu} Zhu, M. et al., "A review of the application of machine learning in water quality evaluation,". Eco Environ. Health \textbf{1}(2), 107–116  (2022).

\bibitem{Ibrahim} Ibrahim, M., Faridah, O., Ibrahim, A. I. N., Alaa-Eldin, M. E. \& Yunus, R. M.,  "Assessment of water quality parameters using multivariate
analysis for Klang River basin," Malays. Environ. Monit. Assess. 187(1), 4182 (2014).

\bibitem{Lopez}  Lopez, E., Schuhmacher, M. \& Domingo, J. L., " Human health risks of petroleum-contaminated groundwater," Environ. Sci. Pollut.
Res. 15(3), 278–288 (2008).

\bibitem{Maione} Maione, C., Nelson, D. R. \& Barbosa, R. M., "Research on social data by means of cluster analysis,"  Appl. Comput. Inform. 15(2),
153–162 (2008).
\bibitem{1} D. A. Goss and R. W. West, \emph{Introduction to the Optics of the Eye} (Butterworth-Hienemann, 2001). 
\bibitem{2} M. P. Keating, Geometric, \emph{Physical and Visual Optics} (Butterworth-Hienemann, 2002). 
\bibitem{3} W. Tasman and E. A. Jaeger, \emph{Duane's Ophtalmology} (LLW, 2013). 
\bibitem{4} T. Callina and T. P. Reynolds, "Traditional methods for the treatment of presbyopia: spectacles, contact lenses, bifocal contact lenses," Ophthalmol. Clin. North Am. \textbf{19}, 25-33 (2006). 
\bibitem{5} Patent Application WO2006011937A2, ``Fluidic Adaptive Lens''. 
\bibitem{6} N. Hazan, A. Banerjee, H. Kim, and C. Mastrangelo, ``Tunable-focus lens for adaptive eyeglasses," Opt. Express \textbf{25}, 1221-1233 (2017).
\bibitem{7} O. Takayama, F. Minotti, and G. Puentes, ``Tunable Fluidic Lenses with High Dioptric Power,"  OSA Continuum \textbf{1}, 181-192 (2018).
\bibitem{8} G. Puentes, D. Voigt, A. Aiello, and J. P. Woerdman, "Experimental observation of depolarized light scattering," Opt. Lett. \textbf{30}, 3216-3219 (2005). 
\bibitem{9} G. Puentes and F. Minotti, ``Spectral Characterization of Optical Aberrations in Fluidic Lenses,"  Front. Phys. \textbf{11}(1299393), 1-14 (2024). 
\bibitem{10} G. Puentes Invention Disclosure Nr.20170102760. ``Adaptive Fluidic Lenses for Subnormal Vision Segment'' (AR109794B1). 
\bibitem{10b} G. Puentes  ``Comparison between neural network clustering, hierarchical clustering, and k-means clustering;: Applications using fluidic fenses,"  \textbf{33}, Optics Express 28405-284019 (2025).
\bibitem{11} F. Schneider, J. Draheim, R. Kamberger, and U. Wallrabe, ``Process and material properties of polydimethylsiloxane (PDMS) for Optical MEMS,"  Sensor and Actuators A \textbf{151}, 95-99 (2006).
\bibitem{12} M. Vallet, B. Berge, and L. Vovelle, "Electrowetting of water and aqueos solutions on poly-ethilene-terephthalate insulating films," Polymer \textbf{37}, 2465-2470 (1996). 
\bibitem{13} T. Krupenking, S. Yang, and P. Mach, "Tunable liquid microlens," Appl. Phys. Lett. \textbf{82}, 316-318 (2003). 
\bibitem{14} S. Kuiper and B. H. Hendriks, "Variable-focus liquid lens for miniature cameras," Appl. Phys. Lett. \textbf{85}, 1128-1130 (2004). 
\bibitem{15} G. C. Knollman, J. L. Bellini, and J. L. Weaver, "Variable-focus liquid-filled hydroacoustic lens,"  J. Acoust. Soc. Am. \textbf{49}, 253-261 (1971).
\bibitem{16} N. Sigiura and S. Morita, "Variable-docus liquid-filled optics lens," App. Opt. \textbf{32}, 4181-4186 (1993). 
\bibitem{17} D. Y. Zhang, V. Lien, Y. Berdichevsky, J. Choi, and Y. H. Lo, "Fluidic adaptivelens with high focal length tenability," App. Phys. Lett. \textbf{82}, 3171-3172 (2003). 
\bibitem{18}K. H. Joeng, G. L. Liu, N. Chronis, and L. P. Lee, "Tunable microdoublets lens array," Opt. Express \textbf{12}, 2494-2500 (2004). 
\bibitem{19} J. Chen, W. Wang, J. Fang, and K. Varahramyan, "Variable focusing microlens with microfluidic chip," J. Micromech. Microeng. \textbf{14}. 675-680 (2004). 
\bibitem{20} N. Chronis, G. L. Liu, K. H. Jeong, and L. P. Lee, "Tunable liquid filled micro-lens array integrated with microfludidic network," Opt. Express \textbf{11}, 2370-2378 (2003). 
\bibitem{21} P. M. Moran, s. Dharmatilleke, A. H. Khaw, and K. W. Tan, ``Fluid lenses with variable focal length,'' App. Phys. Lett. \textbf{88}, 041120-041123 (2006). 
\bibitem{22} H. Ren and S.-T. Wu, ``Variable-focus liquid lens,'' Opt. Express \textbf{15}, 5931-5936 (2007).
\bibitem{23} N. A. Polson and M. A. Hayes, ``Microfluidics controllling fluids in small places,'' Anal. Chem. \textbf{73}, 312A-319A (2001).
\bibitem{37} M. S. N. Siemon, A. S. N.  Shihavuddin, and G. Ravn-Haren , ``Sequential transfer learning based on hierarchical clustering for improved performance in deep learning based food segmentation,''  Sci. Rep. \textbf{11}, 813-827 (2021).

\end{thebibliography}
\end{document}